\documentclass[aip,amsmath,amssymb,reprint]{revtex4-1}
\usepackage{graphicx}  
\usepackage{dcolumn}   
\usepackage{bm}        
\usepackage{amssymb}   
\graphicspath{}
\usepackage{verbatim}
\usepackage[T2A]{fontenc}
\usepackage[utf8]{inputenc}
\usepackage[russian,english]{babel}

\begin{document}

\date{\today}

\title{QED cascade initiation via reflection of a multipetawatt laser pulse from a self-organized parabolic plasma mirror}
\author{M.\,A. Serebryakov}
\affiliation{Institute of Applied Physics RAS, Nizhny Novgorod, Russia}
\email[Corresponding author: ]{serebryakovma@ipfran.ru}
\author{E.\,N. Nerush}
\affiliation{Institute of Applied Physics RAS, Nizhny Novgorod, Russia}
\affiliation{Lobachevsky State University of Nizhny Novgorod, Nizhny Novgorod, Russia}
\author{L. Ji}
\author{X. Geng}
\affiliation{Shanghai Institute of Optics and Fine Mechanics, Chinese Academy of Sciences, Shanghai 201800, China}
\author{I.\,Yu. Kostyukov}
\affiliation{Institute of Applied Physics RAS, Nizhny Novgorod, Russia}
\affiliation{Lobachevsky State University of Nizhny Novgorod, Nizhny Novgorod, Russia}

\begin{abstract}

The self-sustained or avalanche-type cascade is an intriguing prediction of
strong-field quantum electrodynamics (QED) that has yet to be observed in laboratories.
It is accompanied by the conversion of electromagnetic energy into
gamma photons and electron-positron ($e^-e^+$) pairs, whose number increases exponentially
over time. We investigate a simple configuration to initiate a QED cascades: it is based
on the superposition of an incident multipetawatt laser pulse and its reflection
from a solid target. The incident laser pulse <<deforms>> the initially flat target surface,
creating a parabolic mirror that focuses
the reflected radiation. For the considered setup the threshold laser power is about $7\,\text{PW}$. With a
$27\,\text{PW}$ laser pulse, positron production exhibits clear signatures
of an avalanche-type cascade, including exponential
growth and more than 15 positron generations with similar energy spectra.
Therefore, observing an avalanche-type QED cascade does not require
the use of multiple laser channels with precise spatio-temporal synchronization, as previously supposed.

\end{abstract}

\maketitle

\section{\label{sec:Intro} Introduction}

Thanks to remarkable advancements in laser technology, multipetawatt laser systems have now become
available~\cite{sung20174, liang2020recent}. Moreover, projects of sub-exawatt systems
are underway~\cite{peng2021overview, khazanov2023exawatt}, providing extremely high field strength
in the laser focal spot where QED effects can play a key role~\cite{di2012extremely, gonoskov2022charged, fedotov2023advances}.
Thus the development of lasers opens a path for experimental high-field physics.
One of the most fascinating strong-field QED (SFQED) phenomena is the avalanche-type QED cascade,
which efficiently converts light into matter~\cite{goldreich1969pulsar,bell2008possibility,narozhny2015extreme,kostyukov2024physics}.  

A QED cascade develops as a chain-reaction process when the photon emission
by produced particles and the pair photoproduction by emitted photons
alternate over time so that the number of electron-positron pairs and gamma-quanta grows in time exponentially.
QED cascades can develop in so-called shower-type or avalanche-type regimes.
In the shower-type regime the energy of the cascade particles is gained from
the initial high-energy particle and limited by its energy~\cite{sokolov2010pair,lobet2017generation}.
Such cascades develop when a cosmic ray particle passes the Earth atmosphere~\cite{Stanev2010,gaisser2016cosmic} or
when beam of high-energy particles pass through a high-Z target~\cite{zerby1963studies}.
In this scenario, the source of strong electromagnetic (EM) fields is the
nuclei within atoms and molecules, and the pair photoproduction corresponds
to the Bethe-Heitler process~\cite{bethe1934stopping}.
It should be noted that the shower-type cascade have been studied experimentally in laser-beam interaction
\cite{burke1997positron} and in beam-solid interaction \cite{uggerhoj2005interaction, sarri2015generation}
but typically with very low multiplicity
rate.

Opposing to shower-type cascades, the avalanche-type cascades develop in EM fields that provide efficient acceleration
of electrons and positrons~\cite{bell2008possibility}. The EM field also provides production of new electrons,
positrons and photons. In this case the energy of the cascade particles is limited only by the energy
of the EM field thereby dramatically enhancing the cascade multiplicity in comparison with the shower-type cascades.
It is generally believed that the avalanche-type cascades are responsible for
$e^-e^+$ plasma production at neutron stars~\cite{goldreich1969pulsar,harding2006physics, timokhin2010time, medin2010pair}.
The avalanche-type cascades are not yet observed in the laser field.
It is interesting to note that the cascade may limit the attainable intensity of high power lasers~\cite{fedotov2010limitations}
because of laser field absorption in self-generated $e^- e^+$ plasma \cite{nerush2011laser}. 

Efficient photoproduction requires that the electrons and the positrons
are accelerated as much as $\chi \gtrsim 1$ in order to emit photons with
$\varkappa \gtrsim 1$ ensuring enough probability to produce $e^-e^+$ pairs. Here 
\begin{equation}
\chi=\frac{\hbar}{m^{3}c^{4}}\sqrt{-\left(eF_{\mu\nu}p^{\nu}\right)^{2}},
\end{equation} 
is the quantum parameter for the electrons and the positrons, and
\begin{equation}
\varkappa=\frac{\hbar^3 }{m^{3}c^{4}}\sqrt{-\left(eF_{\mu\nu}k^{\nu}\right)^{2}} .
\end{equation}
is the quantum parameter for photons, with $F_{\mu\nu}$ the electromagnetic
field tensor, $p^{\nu}$ the electron 4-momentum, $c$ the speed of light, $e>0$ and $m$ are the
electron charge value and the electron mass, respectively. 

The development of avalanche-type QED cascades requires extremely strong electromagnetic fields,
and as a result, these cascades have not yet been observed in experiments,
making their initiation a challenging and intriguing problem.
Many numerical models of the avalanche-type QED cascades have been proposed~\cite{goldreich1969pulsar, timokhin2010time, fedotov2010limitations,
Elkina2011, nerush2011analytical, nerush2011laser, bashmakov2014effect, gonoskov2015extended, gelfer2015optimized, grismayer2016laser,
grismayer2017seeded, kostyukov2018growth, Bashinov2018, samsonov2019laser, samsonov2021hydrodynamical, mercuri2024growth},
along with various field configurations to facilitate their initiation:
a tightly focused laser pulse \cite{mironov2021onset}, standing waves formed by two
counterpropagating laser beams with different polarizations~\cite{bell2008possibility, kirk2009pair,
Elkina2011, bashmakov2014effect, mironov2014collapse, Grismayer2017, jirka2016electron},
multi-beam configurations~\cite{bulanov2010multiple, gelfer2015optimized, marklund2023towards},
a dipole wave~\cite{gonoskov2012dipole, gonoskov2013probing, magnusson2019multiple},
the field structure of laser-solid interactions~\cite{ridgers2012dense,
bashinov2013electrodynamic, kostyukov2016production, samsonov2019laser,
slade2019identifying, vincenti2019achieving, vincenti2023plasma, Serebryakov2023}.
In addition to the field configuration, the cascade initiation by seed particles
or with various targets is one more important issue for the cascade
development~\cite{artemenko2017ionization, tamburini2017laser, jirka2017qed,
zhu2016dense, guo2024identifying, yitong2021upper}.
It follows from the simulations that the cascade particle dynamics is very
complex~\cite{esirkepov2014attractors,bulanov2017charged,gonoskov2022charged}
and is influenced by many effects like the ponderomotive scattering~\cite{artemenko2017ionization,
gonoskov2017ultrabright, gonoskov2022charged},
the radiative trapping~\cite{ji2014radiation, gonoskov2014anomalous},
QED processes~\cite{di2012extremely, blackburn2020radiation, gonoskov2022charged, fedotov2023advances},
ionization~\cite{artemenko2016formation, artemenko2017ionization, tamburini2017laser} etc.
Moreover the production of $e^- e^+$ plasma is accompanied by collective effects leading
to formation of intricate field-plasma structures~\cite{nerush2011laser,
ridgers2012dense, grismayer2016laser, efimenko2018extreme, yu2018qed, luo2018qed, samsonov2019laser, efimenko2019laser}.

The coherent addition of several laser beams is a direct way to reduce the laser power needed for avalanche-type
cascades~\cite{bulanov2010multiple, gelfer2015optimized, marklund2023towards}. The simulations show that the total laser power can be as low
as $7.2~\text{PW}$ for locally confined cascade in the dipole wave configuration~\cite{gonoskov2017ultrabright}. At the same time, the multi-beam configurations requires several laser channels and their precise spatio-temporal synchronization with phase conjugation that makes the potential experiments difficult since the synchronization of high-power beams is still a challenge \cite{khazanov2023exawatt,nerush2023effect}. For the standing wave configuration formed by two laser beams, the phase conjugation is not necessary but the spatio-temporal synchronization is still needed.
Such synchronization can be avoided if the standing wave is generated by the superposition of the incident laser radiation and that reflected by a plasma mirror.  

Plasma mirror technology has been studied extensively for decades \cite{dromey2004plasma,doumy2004complete,wilson2016ellipsoidal,ouille2024lightwave}. Recently, it has attracted significant attention due to its diverse and high-impact applications, such as enhancing laser radiation contrast \cite{levy2007double,arikawa2016ultrahigh,choi2020highly}, generating high harmonics and ultra-strong electromagnetic fields \cite{gordienko2005coherent,fedeli2021probing,kim2023high,jeong2025toward}, and enabling advanced charged particle acceleration \cite{thevenet2016vacuum,zingale2021emittance,geng2024compact}. Additionally, it plays a key role in developing compact Compton radiation sources \cite{tsai2015compact} and innovative techniques like flying focus \cite{jeong2021relativistic} and relativistic parabolic plasma mirror \cite{jeong2025toward}, thus plasma mirrors demonstrate growing importance in high-intensity laser physics and next-generation light sources.

In this paper we investigate the dynamics of QED cascades in the standing wave formed by the sum of the incident and the reflected
laser radiation. Using the plasma mirror significantly facilitate the setting up the standing wave.
Moreover, the incident laser pulse deforms the initially flat surface of the mirror, thereby leads to focusing of the reflected radiation
and increases the laser field in the focal area~\cite{vincenti2019achieving, vincenti2023plasma}.
The QED effects at laser-solid interactions has been actively explored recently. However, most works either
do not examine the cascade development~\cite{vincenti2019achieving, vincenti2023plasma},
or are limited to the extremely high intensity of the laser field~\cite{kostyukov2016production, samsonov2019laser, Serebryakov2023},
or explore two-pulse configuration forming standing wave~\cite{jirka2017qed,
bashinov2013electrodynamic,  jirka2017qed,  slade2019identifying}. 
In the recent papers the QED cascades are studied in one-pulse configuration with initially non-flat plasma mirrors.
E.g., the cascades development in the field of the incident radiation and the radiation which is reflected and
focused by a conical plasma mirror has been studied~\cite{Samsonov2025production}.   
Here we consider initially flat solid target and laser intensities near the cascade threshold
that is close to the intensity of modern 10-PW laser systems. 

The paper is organized as follows. In Section~II, the configuration of the laser-plasma interaction
is described, and the results of QED particle-in-cell (PIC) simulations are presented. The
growth rate of the positron number is compared with that for cascades in circularly and
linearly polarized standing waves.
In Sec.~III, the QED cascade is considered in detail. Evidence that the cascade
is of the avalanche type
is collected and analyzed. Special attention is paid to positrons of different generations.
The conclusions are given in Sec.~IV.

\section{Field reflection and positron production}

\begin{figure*}
	\includegraphics[width=1\linewidth]{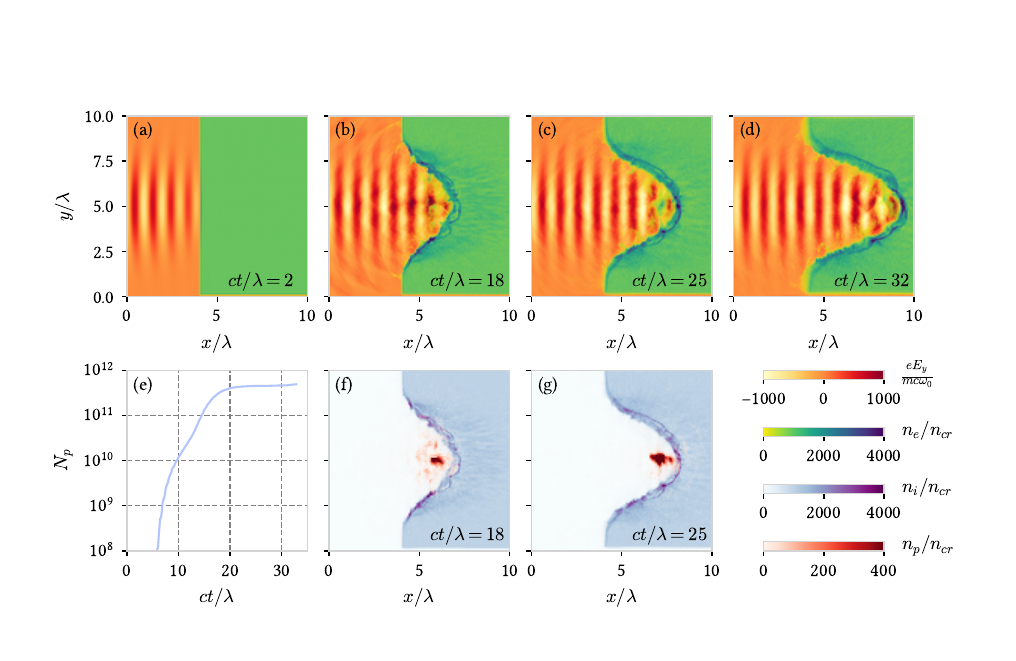}
    \caption{\label{fig1} (a)-(d) and (f)-(g) 3D PIC simulation snapshots for fine-step setup with $a_0 = 700$. The laser field is emitting
    from $x = 0$ box
    boundary. The field reflected by the plasma goes through this boundary thanks to open boundary conditions. (e) Overall number of
    positrons generated in the simulation. }
\end{figure*}

\begin{figure*}
	\includegraphics[width=1\linewidth]{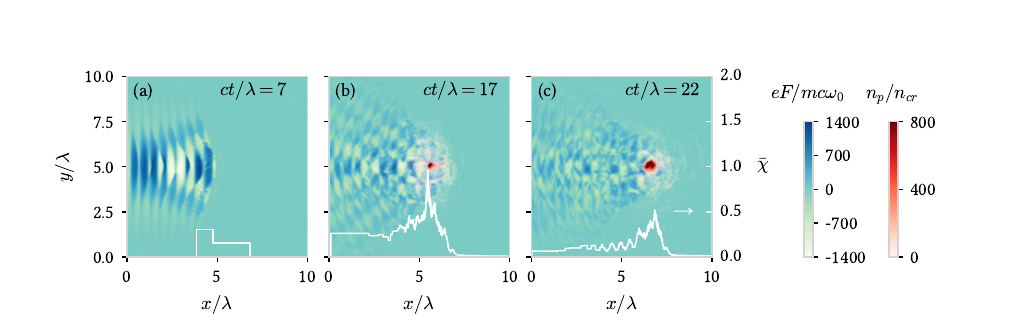}
    \caption{\label{fig2} Field invariant $F = \operatorname{sign}(E^2 - B^2) |E^2 - B^2|^{1/2}$ and positron density from a 3D PIC
    simulation using a fine-step setup with
    $a_0 = 700$. The average quantum parameter $\chi$ of positrons is depicted in white stepped line, with each step representing 400 simulation
    particles along the $x$ axis. }
\end{figure*}

Using reflection, a standing wave can be created with a single laser pulse. For the same amplitude of the standing wave, the energy needed is
twice lower for incident and reflected laser pulse in comparison with two counterpropagating laser pulses. However, reflection of a linearly
polarized petawatt
laser field by a plasma is far from ideal. First of all, the plasma surface oscillates with optical period timescale that leads to copious high
harmonics in the reflected field~\cite{pukhov2010enhanced}. Also, the plasma surface bends due to ion motion;
a pit hole in the plasma is produced which focuses laser pulse like a parabolic
mirror. Then the pit hole elongates and becomes a plasma channel.

Plasma channel formation and focusing of the incident laser pulse are evident in numerical simulations. We simulate the interaction of
mulitipetawatt laser pulse with flat solid target with 3D QED-PIC code QUILL~\cite{quill}. In the modelling, the laser pulse
propagates along the $x$ axis in the simulation box with dimensions $L_x \times L_y \times L_y = 16 \times 10 \times 10 ~ \lambda^3$
with $\lambda = 1\,\mu\text{m}$ the laser pulse wavelength.
The box boundary at $x = 0$ emits this linearly polarized laser pulse which is longer than the simulation box and can't
be placed into the box initially.
The electromagnetic field evolution is computed with FDTD algorithm as usual in PIC simulations.
In paraxial approximation the emitted pulse is a Gaussian beam
focused to a point located $2\,\mu\text{m}$ beneath the target surface.
At the focal plane the spot size is $w_0 = 2 \; \mu\text{m}$ and FWHM duration is $t_f = 120 \; \text{fs}$, and the electric field
has a flattop temporal profile
\begin{multline}
    E_y(y,z,t) = E_0 \cos[\omega_0 (t-t_0)] \exp \left(-\frac{y^2+z^2}{w_0^2}\right) \times \\
	\times
	\begin{cases}  1, & \text{if } |t - t_0| < 0.8 \times t_h \\
	               \frac{1}{2} - \frac{1}{2} \tanh \left( \frac{2 \tau }{1 - \tau^2} \right), & \text{if } 0.8 \times t_h \leqslant |t-t_0| \leqslant 1.2 \times t_h \\
				   0, & \text{otherwise}
    \end{cases}
\end{multline}
with $t_h = t_f/2$, $\tau = (|t - t_0| - t_h) / (0.2 \times t_h)$ and $t_0 = 75\,\text{fs}$ the time
defining the initial distance $ct_0$ between
the laser pulse center and the focusing point.
It is convenient to introduce the dimensionless amplitude of the laser field
$a_0= e E_0/mc\omega_0$, where $E_0$ and $\omega_0 = 2 \pi c / \lambda$ are the
laser field strength and the laser frequency, respectively.
 
The target is a plasma slab with plain surface at $x = 4 ~\mu \text{m}$ and the initial density $n_e = 1000~n_{cr}$,
where $n_{cr} = m\omega_0^2/4\pi e^2$ is the critical plasma density. 
Some computations are performed with 8 simulation electrons and 8 ions per cell and
the following fine-steps: time step $\Delta t = 0.02 / \omega_0$,
and spatial steps $\Delta x = 0.012 \lambda$, $\Delta y = \Delta z = 0.03 \lambda$. Some
other simulations use one electron and one ion per cell and rough steps: time step $\Delta t = 0.339 / \omega_0$,
and spatial steps $\Delta x = 0.025 \lambda$, $\Delta y = \Delta z = 0.1 \lambda$. The rough-step simulations
demonstrate just slightly different laser-plasma dynamics and a bit later cascade onset.

The results of fine-step simulation for $a_0 = 700$ (the laser pulse energy of $3200$\,J) are shown in figures~\ref{fig1} and~\ref{fig2}.
At the initial stage of the interaction the surface of the target is flat.
Then the main part of the laser pulse reaches the target, and the
target surface moves because of the light pressure. This results in the formation of a parabolic-like plasma mirror,
whose focusing point is close
to the plasma surface. The waist in this new focus is less than the initial spot size.
The amplitude of the standing wave in new focal spot is high enough for the cascade onset.
Electrons and positrons in the avalanche are produced near the focus of the plasma mirror.
From $t \approx 8 \lambda / c$ to $t
\approx 18 \lambda / c$ the number of positrons grows exponentially,
see figure~\ref{fig1}~(e). Then the positron production becomes less
efficient. The produced positrons and electrons are trapped near the focus at least
for several laser periods, see figures~\ref{fig1}~(c), (f), (g).

The field reflection is far from ideal. First, the linearly polarization of the laser pulse
forces the plasma surface to oscillate that leads to high
harmonics in the reflected signal~\cite{pukhov2010enhanced}. Second, the mirror shape is not actually parabolic and has a lot of
defects. Third, there is a substantial angle between the propagation direction of the incident and
the reflected fields aside of the focal spot. All this makes the resulting standing
wave quite irregular, see figure~\ref{fig1} and figure~\ref{fig2}. 

The field invariant $F = \operatorname{sign}(E^2 - B^2) |E^2 - B^2|^{1/2}$ is important for QED cascades. If $F < 0$, the field is
mostly magnetic and it's likely that the self-sustained (avalanche-like) cascade can't develop
because secondary electrons and positrons can't gain energy.
If $F > 0$, the field is mostly electric and avalanche-like cascade is more likely.
The regions of high values of $F$ are clearly seen in figure~\ref{fig2}, however
they are not regular and become stochastic especially at later stages.
Average parameter $\chi$ of the positrons is depicted in white stepped line in figure~\ref{fig2}. Every step of the line
covers 400 simulation particles sorted along the $x$ axis. As expected, $\chi$ is higher near the focal spot, and $\chi$ decreases
with time, that is a probable cause of the cascade slowdown for $t \gtrsim 20 \lambda/c$
despite the normalized amplitude of the electric field remains greater than $10^3$.
The electron-positron plasma is compact and its density
reaches relativistic critical density $n_p / n_{cr} \gtrsim a_0$ [see figures~\ref{fig2}(b), (c)].
Plasma probably starts to scatter and absorb the electromagnetic field.
The overall number of produced positrons is about $4.9 \times 10^{11}$.

\begin{figure}
	\includegraphics[width=1\linewidth]{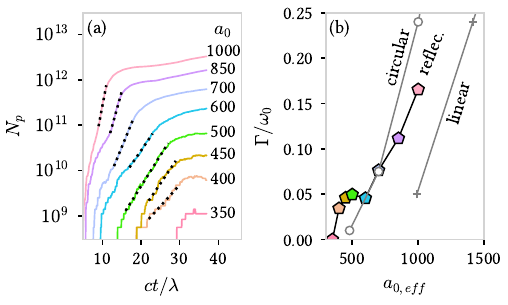}
    \caption{\label{nvst} (a) Number of positrons as function of time in 3D PIC rough-step simulations for $a_0$ values indicated by the labels.
    Intervals exhibiting
    exponential
    growth are fitted with dashed lines (straight lines in the given axes). (b) Growth rates computed from the fits (pentagons) compared
    with values for circularly and linearly polarized standing waves\cite{Grismayer2017} (circles and crosses, respectively).}
\end{figure}

Despite the complicated field structure, the cascade growth rate is surprisingly high.
Rough-step simulations for different field amplitudes ($a_0$ from 350 to 1000) demonstrate
exponential growth of the number of positrons [figure~\ref{nvst}(a)]. Linear fit of PIC data in the logarithmic scale
allows to compute the growth rate $\Gamma$, such that $N_p(t) \propto \exp(\Gamma t)$.
In figure~\ref{nvst}(b) the computed growth rates for cascades in the reflected field
are compared with the growth rates in standing waves of the same energy
formed by two counter propagating laser pulses.
Thus, $a_{0,eff}$ is the amplitude of a single linearly polarized laser pulse which
contains the same energy as the corresponding standing wave. For instance,
if $a_0$ is the amplitude of a single circularly polarized laser pulse, and two such pulses form
a circular standing wave, then $a_{0,eff} = 2 a_0$. In the case of two linearly polarized
laser pulses $a_{0,eff} = \sqrt{2} a_0$.
The data for circularly and linearly polarized standing waves are taken from Ref.~\cite{Grismayer2017}.

The growth rate in sum of incident and reflected fields can be even higher than that in
the rotating electric field or in the circularly polarized standing wave.
High values of the growth rate in the former case can be explained by the
focusing of the field by the plasma
mirror. As the mirror curvature depends on time and the incident field amplitude,
optimal conditions for QED cascades are reached in different time intervals
for different values of $a_0$, see figure~\ref{nvst}(a).
For low laser amplitudes the cascade doesn't start until the mirror curvature reaches some threshold value:
as lower is the field amplitude as later the positrons appear.
For $a_0 = 350$ pair production starts only at $t \approx 30 \lambda / c \approx 100\,\text{fs}$. Thus, the threshold
$a_0$ value for QED cascade onset can strongly depend on the laser pulse duration.
Note also that the time interval where cascade develops can be quite short,
thus the overall number of positrons in the case of reflection
can be lower than in standing waves formed by counter-propagating laser pulses,
even if $\Gamma$ is greater in the former case.

\section{QED cascade details}

\begin{figure*}
	\includegraphics[width=1\linewidth]{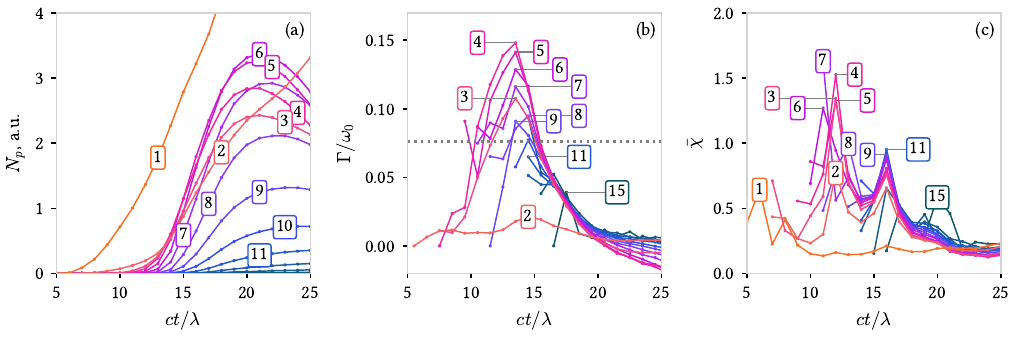}
	\caption{\label{fig3} Positrons of different generations in the fine-step setup with normalized vector potential $a_0 = 700$.
	(a) Number of positrons in each generation which generation numbers indicated by the labels.
	(b) Growth rates computed using equation~(\ref{Gamma_i}) from the results of PIC simulation.
	(c) Average quantum parameter $\chi$ for positrons of different generations. Generation numbers
	are indicated by the labels.}
\end{figure*}

In the considered setup positrons potentially can be produced in three different scenarios: one-step process,
shower-type cascade and avalanche-type cascade. In one-step process initial plasma electrons emit
high-energy photons which produce electron-positron pairs.
In shower-type cascade electrons and positrons, produced in the first step, 
have enough energy to emit subsequent high-energy photons which produce the second generation
of electron-positron pairs. Only in avalanche-type cascades electrons and positrons recover
their energy after the radiation losses, and the newly born electrons and positrons
can also be accelerated. However, in all three scenarios a vast number of positrons
can be produced. Moreover, even in one-step process the number of positrons
can depend on time exponentially as a result of the field increase and sharp
dependence of the photoproduction probability on the field near the threshold.
Thus, the question is which scenario of pair production is realized in practice.

To prove the avalanche-type of QED cascades, the QUILL code was modified so that the
simulation particles have an additional parameter --- the generation number. The initial
plasma electrons belong to the zeroth generation. At each step of the photoproduction,
newly born electrons and positrons get generation
number by one more than that of the parent particles. More than 15 generations
of positrons are produced in the fine-step setup with $a_0 = 700$. The number of
positrons in every generation is shown in figure~\ref{fig3}(a).
In the first generation number of positrons $N_1$ is higher than that in any
other generation. However, sum number of positrons is several times higher
than $N_1$. Thus, definitely multi-step
positron production occurs.

In idealized avalanche-type cascade the number of positrons
in different generations can be described in the following simple model.
Assuming constant source of high-energy photons producing positrons of the first generation,
for the number of such positrons $N_1$ we get
\begin{equation}
\frac{dN_1}{dt} = S,
\end{equation}
with $S$ the source strength. Assuming the same distribution of $\chi$
in all the positron generations, for generation numbers $i \geqslant 2$ we get
\begin{equation}
\label{dNidt}
\frac{dN_i}{dt} = W N_{i-1},
\end{equation}
with $W$ the overall probability of photon emission and subsequent pair photoproduction,
$N_i$ the number of positrons in the $i$-th generation.

The model has a straightforward solution
similar to the Poisson distribution
\begin{eqnarray}
N_1 = t S, & \\
N_i = S W^{i-1} t^i/i!, & \quad i \geqslant 2,
\end{eqnarray}
and the overall number of positrons is
\begin{equation}
N_p = \sum_{i=1}^\infty N_i = \frac{S}{W} \left( e^{Wt} - 1 \right).
\end{equation}
Thus, for $t \gg 1/W$ the cascade growth rate is $\Gamma \approx W$. In practice,
even if the cascade is far from ideal, the growth
rate of every separate generation can be introduced similarly to equation~(\ref{dNidt})
as $\Gamma_i = (1/N_{i-1}) dN_i/dt$. Midpoint approximation for this formula is used
to compute $\Gamma_i$ from the results of PIC simulations
\begin{equation}
\label{Gamma_i}
\Gamma_i(t) \approx \frac{2}{T} \frac{\left[ N_i(t+T/2) - N_i(t-T/2) \right]}{\left[ N_{i-1}(t+T/2) + N_{i-1}(t-T/2) \right]}.
\end{equation}
In figure~\ref{fig3}(b) the growth rates $\Gamma_i$ computed this way are shown for the fine-step simulation with $a_0 = 700$.

The simple model does not take into account the time dependence of $S$ and $W$, as
well as escape of particles from the cascade region. However,
the growth rates computed with equation~(\ref{Gamma_i}) are of the
same order for generations $i \geqslant 3$ and are of the same order
as the overall growth rate $\Gamma$ computed from the slope of overall $N_p(t)$
on the interval $ct/\lambda \in [13, 18]$ and depicted with dotted
line in figure~\ref{fig3}(b).

The average quantum parameter $\bar \chi$ of positrons is about unity,
hence the cascade develop near the threshold of pair photoproduction.
For shower cascade this means a very sharp drop in the growth rate
with the generation number. However, this is not the case
for the results of PIC simulations, for instance
even for $i = 9$ the growth rate remains the same as for lower
generation numbers [see figure~\ref{fig3}(b)]. The average quantum parameter
$\bar \chi$ also doesn't drop with the number of generation. When the cascade
develops, $\bar \chi$ for positrons of every generation remains of the order of $1$,
as shown in figure~\ref{fig3}(c). Only self-sustained avalanche-type cascade
can provide such a picture.
 
\begin{figure}
	\includegraphics[width=1\linewidth]{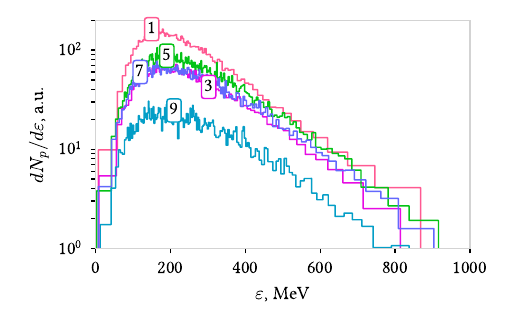}
    \caption{\label{spectrum} Distribution of positron energy in selected generations,
	for the fine-step simulation with $a_0 = 700$ at $t = 18 \lambda / c$. Generation numbers
    are indicated by the labels.}
\end{figure}

\begin{figure}
	\includegraphics[width=1\linewidth]{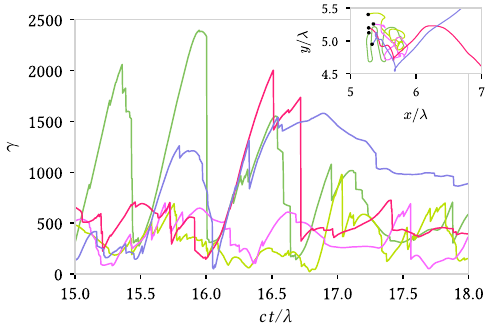}
    \caption{\label{tracks_xy} Lorentz factor for test positrons as a function of time
	for the fine-step simulation with $a_0$ = 700. Inset: trajectories of the same
	test positrons for $t$ ranging from $15 \lambda/c$ (black dots) to $t = 18 \lambda/c$.}
\end{figure}

The main difference between shower- and avalanche-type cascades is that in a shower-type cascade charged particles
are not accelerated by the field, whereas in avalanche-type cascade energy gain is crucial.
Figure~\ref{spectrum} shows positron energy spectrum for different generations at $t = 18 \lambda/c$
for the fine-step setup with $a_0 = 700$. All the spectra extend up to GeV energy range and have almost the same shape.
This is a characteristic feature of self-sustained cascades: stochastic radiation losses and acceleration in
the field lead to an equilibrium distribution function~\cite{nerush2011analytical}, the same for all the generation numbers.
For some test positrons processes of acceleration and photon emission are seen in
figure \ref{tracks_xy}, which shows the dependence of the Lorentz factor of the test positrons on time.
A positron can be smoothly accelerated or decelerated by the field, or can abruptly loose its
energy because of photon emission. Then the positron can be accelerated again.
Subplot demonstrates trajectories of the same positrons in the $xy$ plane
for $t \in [15 \lambda/c, 18 \lambda/c]$, with black dots
depicting start positron positions.
Some positrons are trapped near
the focusing point, and the others can overtake and leave it.

\section{Conclusions}

Avalanche-type QED cascades are typically studied in standing wave setups,
with a preferred field configuration being a rotating electric field at the magnetic node
of a circularly polarized standing wave. Generating such a wave requires
precise alignment and sub-wavelength synchronization of four
linearly polarized laser pulses. In this work we explore a significantly simpler configuration:
a single linearly polarized laser pulse reflected by a plasma slab,
where the QED cascade develops
in the sum field of the incident and reflected waves. PIC
simulations demonstrate that the growth rate of positron number in this
plasma slab setup can be higher than or comparable to that of
a circularly-polarized standing wave of equivalent power.
The high growth rate can be attributed to caused by
ion motion formation of a parabolic-like
mirror from the plasma surface. 
This parabolic mirror effectively focuses the reflected signal,
leading to a reflected amplitude that can noticeably exceed that of the incident pulse.

The evolution of the plasma surface creates optimal conditions
for QED cascades over a short period of time.
For example, a $120\,\text{fs}$,
$27\,\text{PW}$ laser pulse ($a_0 = 700$) interacting with a plasma of initial density
$n_e = 1000 n_{cr}$ results in exponential growth of the positron number
for merely $15\,\text{fs}$.
A long duration is required for formation of
the parabolic shape of the plasma surface. Thus, for a $7\,\text{PW}$
laser pulse positron production starts only after about $100\,\text{fs}$
of interaction. Despite this delay, low power threshold and
simplicity of the setup offer promising prospects for QED cascade
experiments.

A fine-step PIC simulation for a laser pulse with $a_0 = 700$ demonstrates production
of more than 15 positron generations, confirming the cascade mechanism of
positron production. Moreover, the cascade is self-sustaining (avalanche-type)
and the generated electrons and positrons gain energy and the quantum parameter $\chi$ from
the field. As a result, the energy spectra of positrons of different generations
are nearly identical, and almost all positron generations exhibit the
same average quantum parameter $\chi$. Only an avalanche-type cascade
can account such a picture.

Therefore, in the sum field of the incident laser pulse and the pulse reflected
by a plasma slab, QED cascades can develop leading to the production of
approximately $10^9$ or $5 \times 10^{11}$ positrons for laser power of
$7$ and $27\,\text{PW}$, respectively. Additional simulations indicate that
even more favorable conditions for positron production can be found, which will be
the subject of future studies.
Consequently, a single laser channel and
a dense plasma target are sufficient to observe QED cascades,
significantly simplifying the experimental setup.

The research is supported by the Russian Science Foundation (Grant No. 25-12-00336).

\bibliographystyle{unsrt}
\bibliography{plasma-mirror28}

\begin{thebibliography}{10}

\bibitem{sung20174}
Jae~Hee Sung, Hwang~Woon Lee, Je~Yoon Yoo, Jin~Woo Yoon, Chang~Won Lee,
  Jeong~Moon Yang, Yeon~Joo Son, Yong~Ha Jang, Seong~Ku Lee, and Chang~Hee Nam.
\newblock 4.2 pw, 20 fs ti: sapphire laser at 0.1 hz.
\newblock {\em Optics letters}, 42(11):2058--2061, 2017.

\bibitem{liang2020recent}
Xiaoyan Liang, Yuxin Leng, Ruxin Li, and Zhizhan Xu.
\newblock Recent progress on the shanghai superintense ultrafast laser facility
  (sulf) at siom.
\newblock {\em High Intensity Lasers and High Field Phenomena}, pages HTh2B--2,
  2020.

\bibitem{peng2021overview}
Yujie Peng, Yi~Xu, Lianghong Yu, Xinliang WANG, Yanyan LI, Xiaoming LU, Cheng
  WANG, Jun LIU, Chengqiang ZHAO, Yanqi LIU, et~al.
\newblock Overview and status of station of extreme light toward 100 pw.
\newblock {\em The Review of Laser Engineering}, 49(2):93, 2021.

\bibitem{khazanov2023exawatt}
Efim Khazanov, Andrey Shaykin, Igor Kostyukov, Vladislav Ginzburg, Ivan Mukhin,
  Ivan Yakovlev, Alexander Soloviev, Ivan Kuznetsov, Sergey Mironov, Artem
  Korzhimanov, et~al.
\newblock exawatt center for extreme light studies.
\newblock {\em High Power Laser Science and Engineering}, 11:e78, 2023.

\bibitem{di2012extremely}
A~Di~Piazza, C~M{\"u}ller, KZ~Hatsagortsyan, and Ch~H Keitel.
\newblock Extremely high-intensity laser interactions with fundamental quantum
  systems.
\newblock {\em Reviews of Modern Physics}, 84(3):1177, 2012.

\bibitem{gonoskov2022charged}
A~Gonoskov, TG~Blackburn, M~Marklund, and SS~Bulanov.
\newblock Charged particle motion and radiation in strong electromagnetic
  fields.
\newblock {\em Reviews of Modern Physics}, 94(4):045001, 2022.

\bibitem{fedotov2023advances}
A~Fedotov, A~Ilderton, F~Karbstein, Ben King, D~Seipt, H~Taya, and Greger
  Torgrimsson.
\newblock Advances in qed with intense background fields.
\newblock {\em Physics Reports}, 1010:1--138, 2023.

\bibitem{goldreich1969pulsar}
Peter Goldreich and William~H Julian.
\newblock Pulsar electrodynamics.
\newblock {\em Astrophysical Journal, vol. 157, p. 869}, 157:869, 1969.

\bibitem{bell2008possibility}
AR~Bell and John~G Kirk.
\newblock Possibility of prolific pair production with high-power lasers.
\newblock {\em Physical review letters}, 101(20):200403, 2008.

\bibitem{narozhny2015extreme}
NB~Narozhny and AM~Fedotov.
\newblock Extreme light physics.
\newblock {\em Contemporary Physics}, 56(3):249--268, 2015.

\bibitem{kostyukov2024physics}
I~Yu Kostyukov.
\newblock Physics of extremely strong electromagnetic field: Status and
  prospects.
\newblock {\em Bulletin of the Lebedev Physics Institute}, 51(8):S653--S680,
  2024.

\bibitem{sokolov2010pair}
Igor~V Sokolov, Natalia~M Naumova, John~A Nees, and Gerard~A Mourou.
\newblock Pair creation in qed-strong pulsed laser fields interacting with
  electron beams.
\newblock {\em Physical review letters}, 105(19):195005, 2010.

\bibitem{lobet2017generation}
Mathieu Lobet, Xavier Davoine, Emmanuel d’Humi{\`e}res, and Laurent
  Gremillet.
\newblock Generation of high-energy electron-positron pairs in the collision of
  a laser-accelerated electron beam with a multipetawatt laser.
\newblock {\em Physical Review Accelerators and Beams}, 20(4):043401, 2017.

\bibitem{Stanev2010}
Todor Stanev.
\newblock {\em Cosmic ray showers}, pages 175--221.
\newblock Springer Berlin Heidelberg, Berlin, Heidelberg, 2010.

\bibitem{gaisser2016cosmic}
Thomas~K Gaisser, Ralph Engel, and Elisa Resconi.
\newblock {\em Cosmic rays and particle physics}.
\newblock Cambridge University Press, 2016.

\bibitem{zerby1963studies}
CD~Zerby and HS~Moran.
\newblock Studies of the longitudinal development of electron—photon cascade
  showers.
\newblock {\em Journal of Applied Physics}, 34(8):2445--2457, 1963.

\bibitem{bethe1934stopping}
Hans Bethe and Walter Heitler.
\newblock On the stopping of fast particles and on the creation of positive
  electrons.
\newblock {\em Proceedings of the Royal Society of London. Series A, Containing
  Papers of a Mathematical and Physical Character}, 146(856):83--112, 1934.

\bibitem{burke1997positron}
DL~Burke, RC~Field, G~Horton-Smith, JE~Spencer, D~Walz, SC~Berridge, WM~Bugg,
  K~Shmakov, AW~Weidemann, C~Bula, et~al.
\newblock Positron production in multiphoton light-by-light scattering.
\newblock {\em Physical Review Letters}, 79(9):1626, 1997.

\bibitem{uggerhoj2005interaction}
Ulrik~I Uggerh{\o}j.
\newblock The interaction of relativistic particles with strong crystalline
  fields.
\newblock {\em Reviews of modern physics}, 77(4):1131--1171, 2005.

\bibitem{sarri2015generation}
Gianluca Sarri, K~Poder, JM~Cole, W~Schumaker, Antonino Di~Piazza, Brian
  Reville, T~Dzelzainis, D~Doria, LA~Gizzi, G~Grittani, et~al.
\newblock Generation of neutral and high-density electron--positron pair
  plasmas in the laboratory.
\newblock {\em Nature communications}, 6(1):6747, 2015.

\bibitem{harding2006physics}
Alice~K Harding and Dong Lai.
\newblock Physics of strongly magnetized neutron stars.
\newblock {\em Reports on Progress in Physics}, 69(9):2631, 2006.

\bibitem{timokhin2010time}
AN~Timokhin.
\newblock Time-dependent pair cascades in magnetospheres of neutron stars--i.
  dynamics of the polar cap cascade with no particle supply from the neutron
  star surface.
\newblock {\em Monthly Notices of the Royal Astronomical Society},
  408(4):2092--2114, 2010.

\bibitem{medin2010pair}
Zach Medin and Dong Lai.
\newblock Pair cascades in the magnetospheres of strongly magnetized neutron
  stars.
\newblock {\em Monthly Notices of the Royal Astronomical Society},
  406(2):1379--1404, 2010.

\bibitem{fedotov2010limitations}
AM~Fedotov, NB~Narozhny, G{\'e}rard Mourou, and Georg Korn.
\newblock Limitations on the attainable intensity of high power lasers.
\newblock {\em Physical review letters}, 105(8):080402, 2010.

\bibitem{nerush2011laser}
EN~Nerush, I~Yu Kostyukov, AM~Fedotov, NB~Narozhny, NV~Elkina, and H~Ruhl.
\newblock Laser field absorption in self-generated electron-positron pair
  plasma.
\newblock {\em Physical review letters}, 106(3):035001, 2011.

\bibitem{Elkina2011}
N.~V. Elkina, A.~M. Fedotov, I.~Yu. Kostyukov, M.~V. Legkov, N.~B. Narozhny,
  E.~N. Nerush, and H.~Ruhl.
\newblock Qed cascades induced by circularly polarized laser fields.
\newblock {\em Physical Review Special Topics - Accelerators and Beams},
  14(5):054401, May 2011.

\bibitem{nerush2011analytical}
EN~Nerush, VF~Bashmakov, and I~Yu Kostyukov.
\newblock Analytical model for electromagnetic cascades in rotating electric
  field.
\newblock {\em Physics of Plasmas}, 18(8), 2011.

\bibitem{bashmakov2014effect}
VF~Bashmakov, EN~Nerush, I~Yu Kostyukov, AM~Fedotov, and NB~Narozhny.
\newblock Effect of laser polarization on quantum electrodynamical cascading.
\newblock {\em Physics of Plasmas}, 21(1), 2014.

\bibitem{gonoskov2015extended}
A~Gonoskov, S~Bastrakov, E~Efimenko, A~Ilderton, M~Marklund, I~Meyerov,
  A~Muraviev, A~Sergeev, I~Surmin, and Erik Wallin.
\newblock Extended particle-in-cell schemes for physics in ultrastrong laser
  fields: Review and developments.
\newblock {\em Physical review E}, 92(2):023305, 2015.

\bibitem{gelfer2015optimized}
EG~Gelfer, AA~Mironov, AM~Fedotov, VF~Bashmakov, EN~Nerush, I~Yu Kostyukov, and
  NB~Narozhny.
\newblock Optimized multibeam configuration for observation of qed cascades.
\newblock {\em Physical Review A}, 92(2):022113, 2015.

\bibitem{grismayer2016laser}
Thomas Grismayer, Marija Vranic, Joana~Luis Martins, RA~Fonseca, and LO~Silva.
\newblock Laser absorption via quantum electrodynamics cascades in counter
  propagating laser pulses.
\newblock {\em Physics of Plasmas}, 23(5), 2016.

\bibitem{grismayer2017seeded}
Thomas Grismayer, Marija Vranic, Joana~L Martins, RA~Fonseca, and Lu{\'\i}s~O
  Silva.
\newblock Seeded qed cascades in counterpropagating laser pulses.
\newblock {\em Physical Review E}, 95(2):023210, 2017.

\bibitem{kostyukov2018growth}
I~Yu Kostyukov, II~Artemenko, and EN~Nerush.
\newblock Growth rate of qed cascades in a rotating electric field.
\newblock {\em Probl. At. Sci. Technol.}, 2018.

\bibitem{Bashinov2018}
A~V Bashinov, P~Kumar, and A~V Kim.
\newblock Quantum electrodynamic cascade structure in a standing linearly
  polarised wave.
\newblock {\em Quantum Electronics}, 48(9):833--842, September 2018.

\bibitem{samsonov2019laser}
AS~Samsonov, EN~Nerush, and I~Yu Kostyukov.
\newblock Laser-driven vacuum breakdown waves.
\newblock {\em Scientific reports}, 9(1):11133, 2019.

\bibitem{samsonov2021hydrodynamical}
AS~Samsonov, I~Yu Kostyukov, and EN~Nerush.
\newblock Hydrodynamical model of qed cascade expansion in an extremely strong
  laser pulse.
\newblock {\em Matter and Radiation at Extremes}, 6(3), 2021.

\bibitem{mercuri2024growth}
A~Mercuri-Baron, AA~Mironov, C~Riconda, A~Grassi, and M~Grech.
\newblock Growth rate of self-sustained qed cascades induced by intense lasers.
\newblock {\em arXiv preprint arXiv:2402.04225}, 2024.

\bibitem{mironov2021onset}
AA~Mironov, EG~Gelfer, and AM~Fedotov.
\newblock Onset of electron-seeded cascades in generic electromagnetic fields.
\newblock {\em Physical Review A}, 104(1):012221, 2021.

\bibitem{kirk2009pair}
John~G Kirk, AR~Bell, and Ioanna Arka.
\newblock Pair production in counter-propagating laser beams.
\newblock {\em Plasma Physics and Controlled Fusion}, 51(8):085008, 2009.

\bibitem{mironov2014collapse}
AA~Mironov, NB~Narozhny, and AM~Fedotov.
\newblock Collapse and revival of electromagnetic cascades in focused intense
  laser pulses.
\newblock {\em Physics Letters A}, 378(44):3254--3257, 2014.

\bibitem{Grismayer2017}
T.~Grismayer, M.~Vranic, J.~L. Martins, R.~A. Fonseca, and L.~O. Silva.
\newblock Seeded qed cascades in counterpropagating laser pulses.
\newblock {\em Physical Review E}, 95(2):023210, February 2017.

\bibitem{jirka2016electron}
M~Jirka, O~Klimo, SV~Bulanov, T~Zh Esirkepov, E~Gelfer, SS~Bulanov, S~Weber,
  and G~Korn.
\newblock Electron dynamics and $\gamma$ and e- e+ production by colliding
  laser pulses.
\newblock {\em Physical Review E}, 93(2):023207, 2016.

\bibitem{bulanov2010multiple}
SS~Bulanov, VD~Mur, NB~Narozhny, J~Nees, and VS~Popov.
\newblock Multiple colliding electromagnetic pulses: A way to lower the
  threshold of e+ e-pair production from vacuum.
\newblock {\em Physical review letters}, 104(22):220404, 2010.

\bibitem{marklund2023towards}
M~Marklund, TG~Blackburn, A~Gonoskov, J~Magnusson, SS~Bulanov, and A~Ilderton.
\newblock Towards critical and supercritical electromagnetic fields.
\newblock {\em High Power Laser Science and Engineering}, 11:e19, 2023.

\bibitem{gonoskov2012dipole}
Ivan Gonoskov, Andrea Aiello, Simon Heugel, and Gerd Leuchs.
\newblock Dipole pulse theory: Maximizing the field amplitude from 4 $\pi$
  focused laser pulses.
\newblock {\em Physical Review A—Atomic, Molecular, and Optical Physics},
  86(5):053836, 2012.

\bibitem{gonoskov2013probing}
Arkady Gonoskov, Ivan Gonoskov, Christopher Harvey, Antony Ilderton, Arkady
  Kim, Mattias Marklund, G{\'e}rard Mourou, and Alexander Sergeev.
\newblock Probing nonperturbative qed with optimally focused laser pulses.
\newblock {\em Physical review letters}, 111(6):060404, 2013.

\bibitem{magnusson2019multiple}
J~Magnusson, A~Gonoskov, M~Marklund, T~Zh Esirkepov, JK~Koga, K~Kondo, M~Kando,
  SV~Bulanov, G~Korn, CGR Geddes, et~al.
\newblock Multiple colliding laser pulses as a basis for studying high-field
  high-energy physics.
\newblock {\em Physical Review A}, 100(6):063404, 2019.

\bibitem{ridgers2012dense}
CP~Ridgers, Christopher~S Brady, R~Duclous, JG~Kirk, K~Bennett, TD~Arber, APL
  Robinson, and AR~Bell.
\newblock Dense electron-positron plasmas and ultraintense $\gamma$ rays from
  laser-irradiated solids.
\newblock {\em Physical review letters}, 108(16):165006, 2012.

\bibitem{bashinov2013electrodynamic}
AV~Bashinov and AV~Kim.
\newblock On the electrodynamic model of ultra-relativistic laser-plasma
  interactions caused by radiation reaction effects.
\newblock {\em Physics of Plasmas}, 20(11), 2013.

\bibitem{kostyukov2016production}
I~Yu Kostyukov and EN~Nerush.
\newblock Production and dynamics of positrons in ultrahigh intensity
  laser-foil interactions.
\newblock {\em Physics of Plasmas}, 23(9), 2016.

\bibitem{slade2019identifying}
Cody Slade-Lowther, Dario Del~Sorbo, and Christopher~Paul Ridgers.
\newblock Identifying the electron--positron cascade regimes in high-intensity
  laser-matter interactions.
\newblock {\em New Journal of Physics}, 21(1):013028, 2019.

\bibitem{vincenti2019achieving}
Henri Vincenti.
\newblock Achieving extreme light intensities using optically curved
  relativistic plasma mirrors.
\newblock {\em Physical review letters}, 123(10):105001, 2019.

\bibitem{vincenti2023plasma}
Henri Vincenti, Thomas Clark, Luca Fedeli, Philippe Martin, Antonin
  Sainte-Marie, and Neil Zaim.
\newblock Plasma mirrors as a path to the schwinger limit: theoretical and
  numerical developments.
\newblock {\em The European Physical Journal Special Topics},
  232(13):2303--2346, 2023.

\bibitem{Serebryakov2023}
M.~A. Serebryakov, A.~S. Samsonov, E.~N. Nerush, and I.~Yu. Kostyukov.
\newblock Abnormal absorption of extremely intense laser pulses in
  relativistically underdense plasmas.
\newblock {\em Physics of Plasmas}, 30(11), November 2023.

\bibitem{artemenko2017ionization}
II~Artemenko and I~Yu Kostyukov.
\newblock Ionization-induced laser-driven qed cascade in noble gases.
\newblock {\em Physical Review A}, 96(3):032106, 2017.

\bibitem{tamburini2017laser}
Matteo Tamburini, Antonino Di~Piazza, and Christoph~H Keitel.
\newblock Laser-pulse-shape control of seeded qed cascades.
\newblock {\em Scientific reports}, 7(1):5694, 2017.

\bibitem{jirka2017qed}
Martin Jirka, Ondrej Klimo, Marija Vranic, Stefan Weber, and Georg Korn.
\newblock Qed cascade with 10 pw-class lasers.
\newblock {\em Scientific Reports}, 7(1):15302, 2017.

\bibitem{zhu2016dense}
Xing-Long Zhu, Tong-Pu Yu, Zheng-Ming Sheng, Yan Yin, Ion Cristian~Edmond
  Turcu, and Alexander Pukhov.
\newblock Dense gev electron--positron pairs generated by lasers in
  near-critical-density plasmas.
\newblock {\em Nature communications}, 7(1):13686, 2016.

\bibitem{guo2024identifying}
Yinlong Guo, Xuesong Geng, Liangliang Ji, Baifei Shen, and Ruxin Li.
\newblock Identifying quantum effects in seeded qed cascades via laser-driven
  residual gas in vacuum.
\newblock {\em Plasma Physics and Controlled Fusion}, 66(5):055012, 2024.

\bibitem{yitong2021upper}
Wu~Yitong, Ji~Liangliang, and Li~Ruxin.
\newblock On the upper limit of laser intensity attainable in nonideal vacuum
  [j].
\newblock {\em Photonics Research}, 9(4):541--547, 2021.

\bibitem{esirkepov2014attractors}
Timur~Zh Esirkepov, Stepan~S Bulanov, James~K Koga, Masaki Kando, Kiminori
  Kondo, Nikolay~N Rosanov, Georg Korn, and Sergei~V Bulanov.
\newblock Attractors and chaos of electron dynamics in electromagnetic standing
  wave.
\newblock {\em arXiv preprint arXiv:1412.6028}, 2014.

\bibitem{bulanov2017charged}
SV~Bulanov, T~Zh Esirkepov, JK~Koga, SS~Bulanov, Z~Gong, XQ~Yan, and M~Kando.
\newblock Charged particle dynamics in multiple colliding electromagnetic
  waves. survey of random walk, l{\'e}vy flights, limit circles, attractors and
  structurally determinate patterns.
\newblock {\em Journal of Plasma Physics}, 83(2):905830202, 2017.

\bibitem{gonoskov2017ultrabright}
Arkady Gonoskov, Alexey Bashinov, S~Bastrakov, E~Efimenko, A~Ilderton, A~Kim,
  M~Marklund, Iosif Meyerov, A~Muraviev, and Alexander Sergeev.
\newblock Ultrabright gev photon source via controlled electromagnetic cascades
  in laser-dipole waves.
\newblock {\em Physical Review X}, 7(4):041003, 2017.

\bibitem{ji2014radiation}
LL~Ji, A~Pukhov, I~Yu Kostyukov, BF~Shen, and K~Akli.
\newblock Radiation-reaction trapping of electrons in extreme laser fields.
\newblock {\em Physical review letters}, 112(14):145003, 2014.

\bibitem{gonoskov2014anomalous}
Arkady Gonoskov, A~Bashinov, Ivan Gonoskov, C~Harvey, Anton Ilderton, A~Kim,
  Mattias Marklund, G~Mourou, and A~Sergeev.
\newblock Anomalous radiative trapping in laser fields of extreme intensity.
\newblock {\em Physical review letters}, 113(1):014801, 2014.

\bibitem{blackburn2020radiation}
TG~Blackburn.
\newblock Radiation reaction in electron--beam interactions with high-intensity
  lasers.
\newblock {\em Reviews of Modern Plasma Physics}, 4(1):5, 2020.

\bibitem{artemenko2016formation}
Ivan~Igorevich Artemenko, Anton~Aleksandrovich Golovanov, I~Yu Kostyukov,
  Tat'yana~Mikhailovna Kukushkina, Vsevolod~Sergeevich Lebedev,
  Evgenii~Nikolaevich Nerush, Aleksandr~Sergeevich Samsonov, and
  Dmitrii~Andreevich Serebryakov.
\newblock Formation and dynamics of a plasma in superstrong laser fields
  including radiative and quantum electrodynamics effects.
\newblock {\em JETP letters}, 104:883--891, 2016.

\bibitem{efimenko2018extreme}
Evgeny~S Efimenko, Aleksei~V Bashinov, Sergei~I Bastrakov, Arkady~A Gonoskov,
  Alexander~A Muraviev, Iosif~B Meyerov, Arkady~V Kim, and Alexander~M Sergeev.
\newblock Extreme plasma states in laser-governed vacuum breakdown.
\newblock {\em Scientific reports}, 8(1):2329, 2018.

\bibitem{yu2018qed}
JY~Yu, T~Yuan, WY~Liu, M~Chen, W~Luo, SM~Weng, and ZM~Sheng.
\newblock Qed effects induced harmonics generation in extreme intense laser
  foil interaction.
\newblock {\em Plasma Physics and Controlled Fusion}, 60(4):044011, 2018.

\bibitem{luo2018qed}
Wen Luo, Wei-Yuan Liu, Tao Yuan, Min Chen, Ji-Ye Yu, Fei-Yu Li, D~Del~Sorbo,
  CP~Ridgers, and Zheng-Ming Sheng.
\newblock Qed cascade saturation in extreme high fields.
\newblock {\em Scientific reports}, 8(1):8400, 2018.

\bibitem{efimenko2019laser}
ES~Efimenko, AV~Bashinov, AA~Gonoskov, SI~Bastrakov, AA~Muraviev, IB~Meyerov,
  AV~Kim, and AM~Sergeev.
\newblock Laser-driven plasma pinching in e- e+ cascade.
\newblock {\em Physical Review E}, 99(3):031201, 2019.

\bibitem{nerush2023effect}
Evgenii~Nikolaevich Nerush, RR~Iligenov, and I~Yu Kostyukov.
\newblock The effect of pulse phases on the development of electromagnetic
  cascades in the field configuration proposed for the xcels facility.
\newblock {\em Bulletin of the Lebedev Physics Institute}, 50(Suppl
  6):S689--S692, 2023.

\bibitem{dromey2004plasma}
B~Dromey, S~Kar, M~Zepf, and PJROSI Foster.
\newblock The plasma mirror—a subpicosecond optical switch for ultrahigh
  power lasers.
\newblock {\em Review of Scientific Instruments}, 75(3):645--649, 2004.

\bibitem{doumy2004complete}
G~Doumy, F~Qu{\'e}r{\'e}, O~Gobert, M~Perdrix, Ph~Martin, P~Audebert,
  JC~Gauthier, J-P Geindre, and T~Wittmann.
\newblock Complete characterization of a plasma mirror for the production of
  high-contrast ultraintense laser pulses.
\newblock {\em Physical Review E}, 69(2):026402, 2004.

\bibitem{wilson2016ellipsoidal}
Robbie Wilson, Martin King, RJ~Gray, DC~Carroll, Rachel~Jane Dance, Chris
  Armstrong, SJ~Hawkes, RJ~Clarke, DJ~Robertson, D~Neely, et~al.
\newblock Ellipsoidal plasma mirror focusing of high power laser pulses to
  ultra-high intensities.
\newblock {\em Physics of Plasmas}, 23(3), 2016.

\bibitem{ouille2024lightwave}
Marie Ouill{\'e}, Jaismeen Kaur, Zhao Cheng, Stefan Haessler, and Rodrigo
  Lopez-Martens.
\newblock Lightwave-controlled relativistic plasma mirrors.
\newblock {\em Optics Letters}, 49(17):4847--4850, 2024.

\bibitem{levy2007double}
Anna L{\'e}vy, Tiberio Ceccotti, Pascal d’Oliveira, Fabrice R{\'e}au, Michel
  Perdrix, Fabien Qu{\'e}r{\'e}, Pascal Monot, Michel Bougeard, Herv{\'e}
  Lagadec, Philippe Martin, et~al.
\newblock Double plasma mirror for ultrahigh temporal contrast ultraintense
  laser pulses.
\newblock {\em Optics letters}, 32(3):310--312, 2007.

\bibitem{arikawa2016ultrahigh}
Yasunobu Arikawa, Sadaoki Kojima, Alessio Morace, Shohei Sakata, Takayuki Gawa,
  Yuki Taguchi, Yuki Abe, Zhe Zhang, Xavier Vaisseau, Seung~Ho Lee, et~al.
\newblock Ultrahigh-contrast kilojoule-class petawatt lfex laser using a plasma
  mirror.
\newblock {\em Applied Optics}, 55(25):6850--6857, 2016.

\bibitem{choi2020highly}
Il~Woo Choi, Cheonha Jeon, Seong~Geun Lee, Seung~Yeon Kim, Tae~Yun Kim, I~Jong
  Kim, Hwang~Woon Lee, Jin Woo~Yoon, Jae~Hee Sung, Seong~Ku Lee, et~al.
\newblock Highly efficient double plasma mirror producing ultrahigh-contrast
  multi-petawatt laser pulses.
\newblock {\em Optics Letters}, 45(23):6342--6345, 2020.

\bibitem{gordienko2005coherent}
S~Gordienko, A~Pukhov, O~Shorokhov, and T~Baeva.
\newblock Coherent focusing of high harmonics: A new way towards the extreme
  intensities.
\newblock {\em Physical review letters}, 94(10):103903, 2005.

\bibitem{fedeli2021probing}
Luca Fedeli, Antonin Sainte-Marie, Neil Za{\"\i}m, Maxence Th{\'e}venet,
  Jean-Luc Vay, Andrew Myers, Fabien Qu{\'e}r{\'e}, and Henri Vincenti.
\newblock Probing strong-field qed with doppler-boosted petawatt-class lasers.
\newblock {\em Physical review letters}, 127(11):114801, 2021.

\bibitem{kim2023high}
Yang~Hwan Kim, Hyeon Kim, Seong~Cheol Park, Yongjin Kwon, Kyunghoon Yeom, Wosik
  Cho, Taeyong Kwon, Hyeok Yun, Jae~Hee Sung, Seong~Ku Lee, et~al.
\newblock High-harmonic generation from a flat liquid-sheet plasma mirror.
\newblock {\em Nature communications}, 14(1):2328, 2023.

\bibitem{jeong2025toward}
Tae~Moon Jeong, Sergei~V Bulanov, Rashid Shaisultanov, Prokopis Hadjisolomou,
  and Timur~Zh Esirkepov.
\newblock Toward superstrong fields with a relativistic curved plasma mirror.
\newblock {\em Physical Review A}, 111(3):032218, 2025.

\bibitem{thevenet2016vacuum}
M~Th{\'e}venet, A~Leblanc, Subhendu Kahaly, H~Vincenti, A~Vernier,
  F~Qu{\'e}r{\'e}, and J{\'e}r{\^o}me Faure.
\newblock Vacuum laser acceleration of relativistic electrons using plasma
  mirror injectors.
\newblock {\em Nature Physics}, 12(4):355--360, 2016.

\bibitem{zingale2021emittance}
A~Zingale, N~Czapla, DM~Nasir, SK~Barber, JH~Bin, AJ~Gonsalves, F~Isono, J~van
  Tilborg, S~Steinke, K~Nakamura, et~al.
\newblock Emittance preserving thin film plasma mirrors for gev scale laser
  plasma accelerators.
\newblock {\em Physical Review Accelerators and Beams}, 24(12):121301, 2021.

\bibitem{geng2024compact}
Xuesong Geng, Tongjun Xu, Lingang Zhang, Igor Kostyukov, Alexander Pukhov,
  Baifei Shen, and Liangliang Ji.
\newblock Compact laser wakefield acceleration toward high energy with
  micro-plasma parabola.
\newblock {\em Matter and Radiation at Extremes}, 9(6), 2024.

\bibitem{tsai2015compact}
Hai-En Tsai, Xiaoming Wang, Joseph~M Shaw, Zhengyan Li, Alexey~V Arefiev,
  Xi~Zhang, Rafal Zgadzaj, Watson Henderson, V~Khudik, G~Shvets, et~al.
\newblock Compact tunable compton x-ray source from laser-plasma accelerator
  and plasma mirror.
\newblock {\em Physics of Plasmas}, 22(2), 2015.

\bibitem{jeong2021relativistic}
Tae~Moon Jeong, Sergei~V Bulanov, Petr Valenta, Georg Korn, Timur~Zh Esirkepov,
  James~K Koga, Alexander~S Pirozhkov, Masaki Kando, and Stepan~S Bulanov.
\newblock Relativistic flying laser focus by a laser-produced parabolic plasma
  mirror.
\newblock {\em Physical Review A}, 104(5):053533, 2021.

\bibitem{Samsonov2025production}
Alexander Samsonov and Alexander Pukhov.
\newblock Production and magnetic self-confinement of $e^-e^+$ plasma by an
  extremely intense laser pulse incident on a structured solid target.
\newblock {\em Matter and Radiation at Extremes}, 10(5), August 2025.

\bibitem{pukhov2010enhanced}
A~Pukhov et~al.
\newblock Enhanced relativistic harmonics by electron nanobunching.
\newblock {\em Physics of Plasmas}, 17(3), 2010.

\bibitem{quill}
QUILL, \url{https://github.com/QUILL-PIC/Quill}.

\end{thebibliography}

\end{document}